\documentclass[twocolumn,aps,prb,showpacs,amsmath,amssymb,psfrac,
superscriptaddress]{revtex4}
\usepackage{amsmath}

\usepackage{epsfig}
\usepackage{graphicx}
\usepackage{dcolumn}
\usepackage{bm}

\renewcommand{\dag}{^{\dagger}}

\def\gl{\lower.35em\hbox{$\stackrel{\textstyle>}{\textstyle<}$}}
\def\gapp{\lower.35em\hbox{$\stackrel{\textstyle>}{\sim}$}}
\def\lapp{\lower.35em\hbox{$\stackrel{\textstyle<}{\sim}$}}

\begin{document}

\title{Boson-Fermion Duality and Metastability in Cuprate Superconductors}
\author{J.\ Ranninger}
\affiliation{Institut N\'eel, CNRS et Universit\'e Joseph Fourier, BP 166, 
38042 Grenoble Cedex 09, France}
\author{T.\ Doma\'nski}
\affiliation{Institute of Physics, Marie Curie-Sk\l odowska University, 
20-031 Lublin, Poland}
\date{\today}

\begin{abstract} 
The intrinsic structural metastability in cuprate  high T$_c$ materials, 
evidenced in a checker-board  domain structure of the CuO$_2$ planes, locally  
breaks translational and rotational symmetry. Dynamical charge - 
deformation fluctuations of such nano-size unidirectional domains, involving 
Cu-O-Cu molecular bonds, result in resonantly fluctuating diamagnetic pairs 
embedded in a correlated Fermi liquid. As a consequence, the single-particle 
spectral properties acquire simultaneously (i) fermionic low energy 
Bogoliubov   branches for propagating Cooper pairs and  (ii) bosonic localized 
glassy structures for tightly bound states of them at high energies. The 
partial localization of the single-particle excitations results in a fractionation 
of the  Fermi surface as the strength of the exchange coupling between 
itinerant fermions and  partially localized fermion pairs increases upon moving 
from the nodal to the anti-nodal point. This is also the reason why, upon hole 
doping, bound fermion pairs predominantly accumulate near the anti-nodal points 
and ultimately condense in an anisotropic fashion, tracking the gap in the single particle spectrum.

\end{abstract}

\maketitle

\section{Introduction.} 
High T$_c$ superconductivity of the cuprates, it is generally agreed 
upon, emerges out of an unconventional normal state. The most remarkable 
signatures of its strange metal behavior are the pseudogap in the density of 
states and the associated to it remnant Bogoliubov modes. Both show up in a
wide temperature regime above T$_c$  in the single-particle excitations, observed
in angle resolved photoemission spectroscopy (ARPES) \cite{expPG}. Novel 
scanning tunneling microscopy are now able to measure  the spatial distribution 
of  quasi-particle excitations on the atomic length scale \cite{Kohsaka-2007,Gomes-2007,McElroy-2003,Valla-2006,Kohsaka-2008} and  
find  intrinsic 
textured electronic structures, ranging over a wide  regime from low doped 
to optimally doped and beyond. The spatial patterns of the single-particle spectral properties indicate an inter-relation between the low frequency Bogoliubov 
modes and their high frequency counterparts, representing localized glassy 
structures.  In this work we show how this feature can be related to a scenario in 
which itinerant fermionic charge carriers scatter in and out of bosonic tightly 
bound pairs of them in which they are momentarily trapped on nano-size 
deformable molecular clusters. The single-particle excitations thus appear as superpositions of itinerant and localized entities.
\begin{figure}[t]
  \begin{center}
    \includegraphics*[width=3in,angle=0]{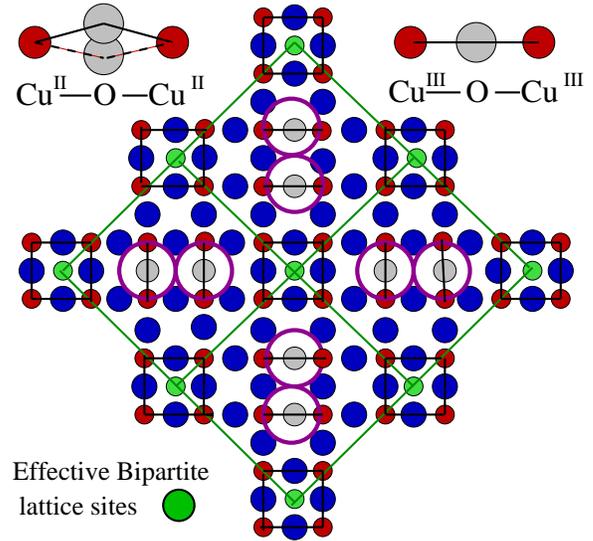}
    \caption{(Color online) An idealized picture of the  local structure of the CuO$_2$ planes, 
    compatible with the STM results \cite{Kohsaka-2007,Gomes-2007,McElroy-2003,Valla-2006,Kohsaka-2008}. 
    It is composed of (i)  
    Cu$_4$O$_{12}$ domains acting as localizing pairing centers with directionally oriented 
    Cu-O-Cu molecular bonds, having central bridging O's (grey circles) which can be displaced out of the 
    CuO$_2$ plane and (ii) Cu$_4$O$_4$ square plaquettes housing the delocalized charge carriers. 
    Small red circles denote Cu cations and the larger blue ones the O anions not directly involved 
    in displacements.}
\label{fig1}
\end{center}
\end{figure}
Ever since the discovery of the high T$_c$ cuprates, experimental evidence for 
their very unusual lattice properties  has become increasingly evident. Apart 
from their well established strongly correlated nature, these compounds 
are metastable single phase materials \cite{Sleight-1991}. Their metastability 
arises from frozen-in structural misfits, involving an incompatibility
between the Cu-O distances of square planar [Cu-O$_4$] configurations in the 
CuO$_2$ planes and of cation-ligand distances in the adjacent layers. 
Metastable compounds have been known for a long time for their intrinsic 
local diamagnetic fluctuations \cite{Vandenberg-1977}, capable of inducing a 
strong pairing component in the many-body ground state wave function. 
The interest in synthesizing materials with such  properties was to bypass the 
stringent conditions on the upper limit of T$_c$, imposed by phonon mediated 
BCS superconductivity \cite{Anderson-1966}.

On a microscopic level, the metastability in the cuprates arises from 
fluctuating [Cu-O-Cu] molecular bonds in the CuO$_2$ planes \cite{Kohsaka-2007,Gomes-2007}. Their deformable ligand 
environments \cite{Cuk-2004,Lee-2006} act as potential pairing centers \cite{McElroy-2003} induced in the undoped systems upon hole doping. These nano 
size domains exhibit an atomic structure \cite{Valla-2006}, which  
locally breaks translational as well as rotational symmetry \cite{Kohsaka-2008}. 
Two degenerate spatially orthogonally oriented [Cu-O-Cu] bonds cause the 
CuO$_2$ plane structure to segregate into a patchwork of orientationally disordered 
domains, separated by a lattice of essentially undeformable molecular 
clusters. Ultimately, this forms an effective bipartite lattice 
structure \cite{Kohsaka-2007,Kohsaka-2008} of the CuO$_2$ planes. The 
charge transfer between the pairing centers and the molecular clusters on 
the lattice surrounding them leads to resonant pairing on that latter. It 
is controlled by  an interplay between localization of the charge 
carriers in form of bound pairs on the pairing centers and their delocalization 
on the lattice which spatially separates them. On a macroscopic level, 
those materials exhibit an overall homogeneous crystal structure 
in a coarse grained sense \cite{Balatsky-2006}. But occasionally, such as in 
La$_{2-x}$Ba$_{x}$CuO$_4$ for x=1/8, the local lattice deformations of the 
pairing centers lock together in a charge ordered phase and thereby impeach superconductivity to occur\cite{Valla-2006}. 
 
\section{THE SCENARIO}
The "formal chemical" Cu valence - not to be confused with its ionic charge -
in the d-hole doped CuO$_2$ planes lies between Cu$^{II}$ and Cu$^{III}$. For 
an isolated undoped CuO$_2$ plane this would correspond to stereochemical 
[Cu$^{II}$-O] distances of 1.94 $\AA$ in the [Cu-O$_4$] basic blocks. The  misfits 
between the atomic structure of the CuO$_2$ planes and those of the adjacent 
layers, which furnish the dopant holes,  push the bridging oxygen of the 
[Cu-O-Cu] bonds out of the CuO$_2$ plane, making them buckled. By doing so, 
they can accommodate the stereochemically assigned inter-atomic distance of 
those bonds.

The scenario for the doped cuprates, which we want to advocate in this work, is 
that the static displacements of the bridging oxygens, which characterize the 
undoped and low doped insulating phase, become dynamic. The fluctuation of 
the bridging oxygens of the [Cu-O-Cu] bonds, in and out of the planes, tends 
to diminish the plane buckling which characterizes the undoped material. 
This tendency gets more and more pronounced  as the doping is increased, driven 
by the increased covalency of the CuO$_2$ basal plane building 
blocks. It however shows a marked slowing down of this behavior as one passes 
through optimal doping\cite{Roehler-2004}. On a microscopic level, this implies 
fluctuations between  kinked [Cu$^{II}$ - O - Cu$^{II}$] molecular 
bonds (characteristic for  the undoped systems) and straightened out ones 
[Cu$^{III}$ - O - Cu$^{III}$], with an ideal stereochemical [Cu$^{III}$-O] distances 
of 1.84 $\AA$. In this process two electrons get momentarily captured in the 
local dynamically deformable structure of the CuO$_2$ planes. It results in 
locally correlated charge-deformation fluctuations which  break up the 
over-all homogeneous structure of the cuprates into a checker-board 
structure, as scanning tunneling microscopy (STM) results  (Figs. 4 and 5 in
Ref. 6) have shown. The net difference in length between the 
two different molecular bonds on such charge-deformation fluctuating
checker-board pairing domains can  be reduced (i) because of the dynamical 
nature of these pairing fluctuations and (ii) because it involves 
cooperatively several of such [Cu-O-Cu] bonds.

The likelihood of a segregation of a homogeneous lattice structure into 
polaronic domains, embedded in a non-polaronic matrix, such as advocated 
in the present scenario, had been speculated upon for a long time. For the case 
of intermediate electron-lattice coupling and the adiabatic to anti-adiabatic 
cross-over regime, individual itinerant charge carriers are known to fluctuate 
in and out of localized polaronic states \cite{Polaronworks}. Unfortunately,
the  present state of art of the theory of many-polaronic systems can still
not handle situations other than for  homogeneous or globally symmetry broken 
solutions. Nevertheless indications for resonant pairing in such systems 
exist,  where the single-particle spectral function has both coherent 
delocalized contributions and localized ones in form of localized polarons, respectively bipolarons. This has been discussed in the framework of dynamical 
mean field theory, numerical renormalization group and Monte Carlo 
studies \cite{DMFT-polarons}.

Given the complexity of the inter-related charge-deformation dynamics in 
such systems, it appeared judicious to introduce a phenomenological  
Boson-Fermion model (BFM), to capture the salient features of such intrinsically 
locally dynamically unstable systems with a tendency to segregate into  
subsystems of localized and itinerant charge  carriers. This idea was 
originally proposed by one of us (JR) in the  early eighties in an attempt 
to describe the abrupt cross-over between a weak coupling adiabatic 
electron-phonon mediated BCS superconductor and an insulating state, 
respectively superconducting phase, of  bipolarons in the strong coupling 
anti-adiabatic regime. The essential features of this conjectured BFM was to 
introduce an effective local boson-fermion exchange coupling between 
polaronically bound pairs and itinerant charge carriers. This picture 
has been substantiated subsequently by small cluster calculations for
electrons strongly coupled to localized lattice vibrational 
modes \cite{Ranninger-2006-2008}. It permits to relate the effective 
boson-fermion exchange coupling back to the parameters, characterizing 
the electron-lattice coupled system, ie., local phonon frequency and 
electron-phonon coupling.

In order to cast into a tractable model the physics of dynamically fluctuating
[Cu-O-Cu] bonds, which trigger local double charge fluctuations, we present
in Fig. 1 an idealized structure for such a local checker-board bipartite 
lattice structure, which comes very close to the actually observed structure. 
The corresponding checker-board pairing centers consist of Cu$_4$O$_{12}$ 
domains (three nearest neighbor Cu-Cu distances across) on which charge 
carriers pair up, driven by polaronic effects. The lattice deformations 
of adjacent Cu$_4$O$_{12}$ domains are assumed to be uncorrelated in order 
to prevent the system to undergo a global lattice instability. The 
orientational randomness of the [Cu-O-Cu] unidirectional bonds, together with 
the quadratic Cu$_4$O$_4$ plaquettes (see Fig.1), which  separate those 
polaronic Cu$_4$O$_{12}$ domains, justifies that. Ultimately, this results  
in the picture of an overall bipartite lattice structure for the CuO$_2$ planes 
with a periodicity of four nearest neighbor Cu-Cu distances. d-holes on the 
non-polaronic Cu$_4$O$_4$ plaquettes in the cuprates are known to behave 
as delocalized, though strongly correlated, entities  subject to 
d$_{x^2 - y^2}$ - wave pairing correlations \cite{Hirsch-1988,Altman-2002}. 
In the present study we shall concentrate on the purely lattice driven 
pairing aspects in the cuprates, caused by their intrinsic metastabilities. We 
hence neglect here any Hubbard type correlations leading to hole pairing and 
treat the Cu$_4$O$_4$ square plaquettes as effective lattice sites on which 
the charge carriers behave as itinerant uncorrelated quasi-particles. 
When they hop on and off the Cu$_4$O$_{12}$ pairing centers, 
they  interact with their local dynamical deformations. The resulting local 
physics for resonant pairing for such a set-up and its manifestations in the 
electronic and phononic spectral properties have been studied in some detail 
by exact diagonalization studies \cite{Ranninger-2006-2008}. 

Indications for resonant pairing in the cuprates, driven by local dynamical 
lattice fluctuations can be found in quite a variety of experimental studies: 
the longitudinal optical (LO) Cu-O bond stretching mode of about 60 meV appears 
strongly coupled to charge carriers 
near the {\it hotspot} anti-nodal points in the Brillouin zone (BZ) $[q_x,q_y]=
[\pm \pi/2,0],[0, \pm \pi/2]$  \cite{Cuk-2004,Lee-2006}. Their pairing results
in the pseudogap feature, setting in when reducing the temperature T to below 
a certain, strongly doping dependent, T$^*$. Upon entering the superconducting 
doping regime, coming from the insulating parent compound, this LO mode 
splits into two modes, separated by $\simeq$ 10 meV \cite{Reznik-2006}. 
This indicates a crystal lattice symmetry breaking, linked to dynamical 
charge inhomogeneities which are absent in the underdoped and overdoped insulating phases. Pressure \cite{Haefliger-2006}, 
isotope substitution studies \cite{Rubio-Temprano-2000} and  atomic resolution 
$d^2I/d V^2$-spectroscopy \cite{Lee-2006} show concomitant anticorrelated modulations
of the pseudo-gap size and the frequency of this LO buckling mode. Correlated
charge-deformation fluctuations, related to a resonant pairing superconducting phase show up in 
the onset of a macroscopic superfluid state of the charge carriers together with 
changes in the local lattice dynamics which acquires phase correlated macroscopic features. 
They are seen in Rutherford back scattering experiments \cite{Sharma-1996}, an abrupt 
decrease in the kinetic energy of  local vibrational modes \cite{Mook-1990}, a similar abrupt 
increase of a low energy electronic background, seen in near IR excited Raman 
scattering \cite{Ruani-1997} and an  increase in intensity of certain Raman active
phonon modes \cite{Misochko-1999}, indicative of changes in the scattering mechanism 
involving the charge carriers and local lattice modes.

\section{THE MODEL}
Superconductivity in the cuprates is  destroyed, exclusively, by phase 
fluctuations of a bosonic order parameter \cite{Emery-Kivelson-1995,Franz-1998},  
with the finite amplitude of it, being already established well above T$_c$. 
It reflects the local nature of the Cooper-pairs, whose signature is (i) a $T_c$ scaling with the zero temperature density of superfluid carriers \cite{Uemura-1989} and (ii) the XY character of the transition \cite{Salamon-1993}.  Going into the normal state, above T$_c$, the propagating 
Cooperons  become diffusive and the superconducting gap changes into a pseudogap 
in a continuous fashion \cite{Devillard-2000}. The observed Nernst \cite{Wang-2006}, transient Meissner effect \cite{Corson-1999} and the proximity induced 
pseudogap \cite{Yuli-2009} bare this out. The  gap in  the single-particle 
spectrum and the  diffusively propagating strongly bound Cooper pairs testify 
the competition between amplitude and phase fluctuations of the order parameter 
in form of an anti-correlated T$_c$ versus T$^*$ variation upon changing 
the hole doping \cite{Tesanovic-2008,Huefner-2008}. The insulating, not 
antiferromagnetically ordered glassy state, at low temperature and low doping 
can be envisaged as a Mott correlation driven state of phase 
uncorrelated singlet-{\it bonding pairs}. With increased doping, this 
insulating state changes into a superconducting phase correlated state of such 
{\it bonding pairs} \cite{Cuoco-2006,Stauber-2007}. 
{\it Bonding pairs} are defined by local linear superpositions of bound pairs 
and  pairs of itinerant charge carriers. To what extent such an insulating state could result from a Cooper-pair Wigner crystallization, has been investigated \cite{Tesanovic-2004,Pereg-Barnea-2006}.  

The features which characterize the normal and superconducting phase of the cuprates 
necessitate to treat amplitude and phase fluctuations on an equal footing. This had 
originally also been the objective in conjecturing the BFM and to project out coexisting 
effective bosonic and fermionic charge excitations for systems which are at the frontier 
between amplitude 
fluctuation driven BCS superconductors and a phase fluctuation driven superfuidity of tightly 
bound real-space pairs. The BFM is designed to treat a \underline{single component} system, 
where at any given moment a certain percentage of the charge carriers is locally paired and 
thus results in a finite bosonic amplitude.  This is achieved by imposing a common chemical 
potential (determined by the bosonic energy level) for the fermionic and bosonic charge carriers. 
A  charge exchange term, linking the fermionic and bosonic subsystem, then controls the 
inter-related dynamics  between amplitude and phase fluctuations. It drives the system either 
to an insulating or superfluid state with corresponding superconducting, respectively 
insulating, gaps being centered at the chemical potential. The opening of such gaps 
does not depend on any particular set of Fermi wavevectors and hence is unrelated to any 
global translational symmetry breaking.

The degree of  anisotropy of pairing and of the charge carrier dispersion 
in the CuO$_2$ planes monitors the relative importance of localization 
versus delocalization  in different regions of the Brillouin zone. Near the
anti-nodal points, strong pairing results from strong intra-{\it bonding pair} correlations between bound pairs on the pairing centers 
and their itinerant counterparts in their immediate vicinity \cite{Ranninger-2006-2008}. It leads to their partial localization, which 
shows up in form of a pseudogap in the single-particle spectral properties 
and  destroys the Fermi surface. As one moves toward the nodal points, 
$[k_x,k_y] = [\pm \pi/2,\pm \pi/2]$, along the socalled Fermi arc in the 
Brillouin zone (corresponding to the Fermi surface in the non-interacting 
system), those intra-{\it bonding pair} phase correlations are weakened. The 
degree of localization then reduces and with it, the size of the pseudogap. 
At the same time, inter-{\it bonding pair} phase correlations between 
neighboring pairing domains come into play and with it, superconducting phase 
locking. At low energies, this leads to Bogoliubov like modes, which emerge out
of localized phase uncorrelated {\it bonding pairs}. We derive below these
properties on the basis of the BFM, adapted to the specific anisotropic 
features of the cuprates. 

Transposing our picture of the cuprate molecular structure (Fig. 1) onto the 
BFM (see also Figs. 3 and 4 in Ref.~\cite{Ranninger-2010}) implies the 
following: We introduce effective lattice sites, which are 
composed of two components: One which represents the pairing centers 
(the Cu$_4$O$_{12}$ domains) and describes selftrapped bosonic pairs of charge 
carriers. The other one which describes the itinerant charge carriers on the
four-site ring, constituted of the Cu$_4$O$_4$ plaquettes, taking into account
that each such plaquette is shared by  four neighboring pairing centers. For 
the undoped  half-filled band situation, with one electron per Cu site, we thus 
have four itinerant electrons on the ring, belonging to a specific pairing
center and four electrons being localized in form of two 
Cu$^{II}$-O-Cu$^{II}$ bonds on the pairing centers. Deviating from the 
undoped limit upon doping n$_h$ holes per Cu ion into the systems, reduces the 
concentration of Cu$^{II}$-O-Cu$^{II}$ bonds in the trapping centers by 
n$_B$  $\simeq$ $\frac{1}{2}$ n$_h$ . This opens up the phase space for 
itinerant electrons from the four-site ring to hop on off those trapping 
centers. Such a resonant scattering process converts a small
number n$_B$ of those itinerant charge carriers 
into bosonic bound pairs. Following the experimental results of the strong 
changes in local lattice properties with hole doping, we assume that hole 
doping monitors exclusively the concentration of the Cu$^{II}$-O-Cu$^{II}$ 
bonds and that hence the total number of itinerant electrons and induced pairs
of them will remain roughly the same as it was in the undoped case,  i.e., 
n$_{tot}$ = n$_F$ + 2n$_B$ = 1.

The d-wave paring symmetry of those systems imposes an analogous d-wave symmetry 
for the exchange interaction between (i) pairs 
of itinerant charge carriers $c^{(\dagger)}_{{\bf k}\sigma}$, corresponding 
to the "plaquette site" states  and (ii) polaronicaly bound  pairs of them 
$b_{\bf q}^{(\dagger)}$, corresponding to  the "pairing center site" states. 
The Hamiltonian describing such a scenario is then given by
\begin{eqnarray}
H_{BFM} &=& H^0_{BFM} + H^{exch}_{BFM} \\
H^0_{BFM} &=& \sum_{{\bf k}\sigma}(\varepsilon_{\bf k} -\mu)c_{{\bf k}\sigma}^{\dagger} 
c_{{\bf k}\sigma}+\sum_{\bf q} 
(E_{\bf q}-2\mu) b_{\bf q}^{\dagger} b_{\bf q}.\qquad\\
H^{exch}_{BFM}&=&(1/\sqrt{N})\sum_{{\bf k},{\bf q}}(g_{{\bf k},{\bf q}}b_{\bf q}\dag 
c_{{\bf q-k},\downarrow}c_{{\bf k},\uparrow}+H.c.),
\label{H_exch}
\end{eqnarray}
The anisotropy, which characterizes the electronic structure of cuprates, is contained 
in the standard expression for the bare charge carrier dispersion given by 
$\varepsilon_{\bf k} = -2t[cos k_x + cos k_y] + 4t'cos k_x cos k_y$ of the 
CuO$_2$ planes with $t'/t=0.4$ and the bare d-wave exchange coupling 
$g_{{\bf k},{\bf q}} = g[cos k_x  - cos k_y]$. Given the polaronic origin of the  
localized pairs of tightly bound charge carriers, we assume them as  dispersionless 
bosonic excitations with $E_{\bf q}=2\Delta$. 

The charge exchange term $H^{exch}_{BFM}$ controls the transfer of electrons 
(holes) between real and momentum space \cite{Hanaguri-2008} and monitors the 
interplay between the delocalizing  and the localizing effect. Depending on the strength of 
the exchange coupling  $g_{{\bf k},{\bf q}}$, it results in a 
competition  between local intra-{\it bonding pair} correlations, favoring insulating features,
and spatial inter-{\it bonding pair} correlations, favoring  superconducting phase locking \cite{Cuoco-2006}. The 
fermionic particles thereby acquire contributions coming from the bosonic particles and the 
bosonic particles having features derived from their fermionic constituents. As we shall see
below, the physically meaningful fermions in such a system are superpositions  of fermions  and 
bosonic bound fermion pairs, accompanied by fermion holes. This boson-fermion duality, which
characterizes the electronic state of the cuprates, results from 
the  "duplicituous"\cite{Hanaguri-2008} nature of their charge carriers, which 
supports simultaneously superconducting correlations
in momentum space (fermionic Bogoliubov excitations) and real space correlations resulting 
in the pseudogap (derived from localized bosonic bound fermion pairs). This apparent "schizophrenic"
behavior \cite{Goss-Levi-2007} of the quasi-particles can be traced back to their different energy 
scales characterizing their excitations. Large excitation energies (above the Fermi energy) 
characterize their localized selftrapped nature and small excitation energies (below the Fermi energy) their 
quasi-coherently propagating Cooper pair nature.

In order to obtain the spectroscopic features of effective fermionic and bosonic excitations we
have to reformulate this interacting Boson-Fermion mixture in terms of two effective 
commuting Hamiltonians, one describing purely fermionic excitations and one purely 
bosonic ones. The boson-fermion interaction thereby is absorbed into inter-dependent
coupling constants by renormalizing $g_{{\bf k},{\bf q}}$ down to zero via a 
flow-equation renormalization approach \cite{Wegner-1994}. At every step of this procedure the 
renormalized Hamiltonian is projected onto the basic structure given by  $H^0_{BFM}$  plus a 
renormalization generated fermion-fermion interactions term\cite{Domanski-2001} 
\begin{eqnarray}
H^{F-F}_{BFM} = {1  \over N} \sum_{{\bf p},{\bf k},{\bf q}} U^{F-F}_{{\bf p},{\bf k},{\bf q}}
c^{\dagger}_{{\bf p} \uparrow}c^{\dagger}_{{\bf k} \downarrow}c_{{\bf q} \downarrow}
c_{{\bf p}+{\bf k}-{\bf q} \uparrow}.
\end{eqnarray}
This is achieved by transforming the Hamiltonian in infinitesimal steps,
controlled by a flow parameter $\ell$ in tems of repeated unitary transformations 
$H(\ell)=e^{S(\ell)}He^{-S(\ell)}$, resulting in  differential equations 
$\partial_\ell H(\ell)=[\eta(\ell),H(\ell)]$ with 
$\eta(\ell) \equiv (\partial_\ell e^{S(\ell)}/\partial_\ell)e^{-S(\ell)}$, determining the flow of the parameters of our system. In its canonical form \cite{Wegner-1994}, $\eta(\ell)=[H^0(\ell),H(\ell)]$ presents an 
anti-Hermitean generator. For details of the ensuing coupled non-linear differential 
equations for the various $\ell$ dependent parameters $\varepsilon_{\bf k}(\ell),
E_{\bf q}(\ell), U^{F-F}_{{\bf p},{\bf k},{\bf q}}(\ell), g_{{\bf k},{\bf q}}(\ell), \mu(\ell)$ 
we refer the reader to our previous work \cite{Domanski-2001,Domanski-2003a}. The parameters,
characterizing $H^0$ and $H^{exch}$, evolve as the flow parameter $\ell$  increases.  
The renormalization procedure starts with  $\ell=0$, for which they are given by the bare values
$\varepsilon_{\bf k}, E_{\bf q} = 2\Delta, g_{{\bf k},{\bf q}}$ together with 
$U^{F-F}_{{\bf p},{\bf k},{\bf q}} \equiv 0$. The chemical potential
$\mu(\ell)$ is chosen at each step of the renormalization flow such as to fix a 
given total number of fermions and bosons. The flow of these parameters converges for  
$\ell \rightarrow \infty$ and results in two uncoupled  systems: one for the effective fermionic 
excitations and one for the effective bosonic ones with a fix point Fermion dispersion 
$\varepsilon^*_{\bf k} = \varepsilon_{\bf k}(\ell \rightarrow \infty)$.
For isotropic exchange coupling and fermion dispersion this problem had been studied 
previously \cite{Domanski-2001,Domanski-2003a,Stauber-2007}, predicting 
the pseudogap \cite{Ranninger-1995} and damped Bogoliubov modes \cite{Domanski-2003a} 
in angle resolved photoemission spectra. Both have since been verified 
experimentally \cite{expPG}.

\section{The Boson-Fermion duality.} 
The anisotropy of the electronic structure of cuprates tracks a change-over 
from self-trapped (localized) fermions, in form of diffusively propagating bosonic pairs, 
into itinerant propagating (delocalized) fermions upon going from the anti-nodal to the nodal point 
on an arc in the Brillouin zone, determined by $\varepsilon^*_{{\bf k}_F}(\phi) = \mu$. 
To illustrate that, we  evaluate the single-particle spectral function for wave vectors 
${\bf k}=|{\bf k}|[sin \phi,cos \phi]$, orthogonally intersecting this arc at various 
${\bf k}_F(\phi)$, where the motion of the charge carriers is essentially one dimensional. $\phi$ 
denotes the angle of those ${\bf k}$-vectors with respect to the line $[\pi,\pi]- [\pi,0]$, (see Fig.3). 

In order to relate our study to a nearly  half filled band situation,
characterizing the doped cuprates, we choose $\Delta \simeq 0.75$ (in units of
a nominal fermionic band width of 8t), with the bosonic level lying just barely 
below the center of the itinerant fermion band such as to reproduce the typical
shape of the CuO$_2$ planar Fermi surface. Our choice of the 
boson-fermion exchange coupling strength g = 0.1, yields a typical
onset temperature T$^* = 0.016$  for the pseudogap of roughly a hundred 
degrees K. For a characteristic temperature of the pseudogap phase
($T= 0.007 < T^*$), it implies a concentration of itinerant fermionic charge carriers $n_F = \sum_{{\bf k}\sigma}\langle c^{\dagger}_{{\bf k}\sigma}c_{{\bf k}\sigma}\rangle = 0.88$ and that of self-trapped ones  bound into fermion pairs,
$n_B = \sum_{\bf q} \langle b^{\dagger}_{\bf q} b_{\bf q}\rangle = 0.075$. 
This corresponds to  a hole doping $n_h=0.12$, with a total number of carriers
of $n_{tot} = n_F + 2n_B =1.03$. Hole doping redistributes the relative
occupation of fermions and bosons which ultimately leads to a shrinking of the 
arcs (see section V). The charge carriers around the nodal point are primarily 
given by delocalized fermionic one-particle states, while at the hotspot 
anti-nodal points they are  localized bosonic bound fermion pairs. Yet, as we shall see below, they  will become itinerant and 
eventually condense as the temperature is decreased. The reason for that is that 
the bare exchange coupling $g_{{\bf k},{\bf q}}$ is equal to zero at the nodal 
point ($\phi = \pi/4$) and increases as one moves to the anti-nodal points 
($\phi = 0, \pi/2$), where it reaches its maximal value, equal to g. As a consequence, $\varepsilon^*_{\bf k}$ remains essentially unrenormalized for
${\bf k}$ vectors crossing the arc near the nodal point. Upon approaching the 
anti-nodal point, on the contrary, $\varepsilon^*_{\bf k}$ acquires a 
sharp S-like inflexion at ${\bf k}_F(\phi)$, which leads to the the appearance
of the pseudogap in the single-particle density of states. 

Our prime objective in the present study is to disentangle the contributions to 
the single-particle spectral function coming from the itinerant and from 
the localized features. The latter arise from single-particles being 
momentarily trapped in form of localized pairs. The effective fermionic and 
bosonic excitations are obtained in a renormalization procedure similar 
to that of the Hamiltonian, but this time by applying it to the fermion and 
boson operators themselves \cite{Domanski-2003a,Domanski-2004}. The evaluation 
of the single-particle spectral function 
\begin{eqnarray}
A^F({\bf k}, \omega) &=& -{1 \over \pi} Im \int_0^{\beta} d \tau G^F({\bf k},\tau)
e^{(\omega +0^+)\tau} \qquad\qquad \nonumber \\
G_F({\bf k},\tau) &=& \langle \langle c_{{\bf k} \uparrow}(\tau);
c^{\dagger}_{{\bf k}\downarrow}\rangle\rangle_H
\end{eqnarray}
in a correspondingly renormalized manner is achieved by applying the unitary 
transformation $e^{S(\ell)}$ to the Green's function itself. It results in
\begin{eqnarray}
\langle \langle c_{{\bf k} \sigma}(\tau);c^{\dagger}_{{\bf k}\sigma}(0)\rangle\rangle_H= 
\qquad\qquad\qquad\qquad\qquad\qquad\quad\nonumber &\\
\langle \langle e^{S(l)}e^{\tau H(\ell)}c_{{\bf k} \sigma}e^{-\tau H(\ell)}e^{-S(l)};e^{S(l)}
c^{\dagger}_{{\bf k}\sigma}e^{-S(l)}\rangle\rangle_{H(\ell)} = \nonumber & \\
\langle \langle  e^{S(\infty)}e^{\tau H^*}c_{{\bf k} \sigma}e^{-\tau H^*}
e^{-S(\infty)};e^{S(\infty)} c^{\dagger}_{{\bf k}\sigma}e^{-S(\infty)}\rangle\rangle_{H^*}, &
\end{eqnarray}
where the trace has to be carried out over the fully renormalized fixed point Hamiltonian $H^*$.
Neglecting the residual interaction $U^{F-F}_{{\bf p},{\bf k},{\bf q}}$ between the fermions 
and restricting ourselves to the pseudogap phase without any long range phase locking, we obtain 
the following renormalized fermion operators \cite{Domanski-2004}:
\begin{eqnarray}
{c^{\dagger}_{-{\bf k},-\sigma}(\ell) \brack c_{{\bf k},\sigma}(\ell)} &=& 
u^F_{\bf k}(\ell) {c^{\dagger}_{-{\bf k},-\sigma} \brack c_{{\bf k},\sigma}} \nonumber \\
&&\mp \frac{1}{\sqrt N}\sum_{\bf q}v^F_{{\bf k},{\bf q}}(\ell){b^{\dagger}_{\bf q} 
c_{{\bf q+k}, \sigma} \brack  b_{\bf q} c^{\dagger}_{{\bf q-k}, -\sigma}}, 
\end{eqnarray}
with $\ell$ dependent parameters $u^F_{\bf k}(\ell), v^F_{\bf k}(\ell)$  determined by the 
flow equations. The single-particle fermionic spectral function resulting from such a procedure
\begin{eqnarray}
A^F({\bf k},\omega) = |u^F_{\bf k}(\infty)|^{2} \delta \left( \omega\!+\!\mu\!-\!\varepsilon^*_{\bf k} 
\right) \qquad \nonumber \\
+ \frac{1}{N} \sum_{{\bf q}\neq{\bf 0}} \left( n_{\bf q}^{B} + n_{{\bf q}-{\bf k}\downarrow}^{F} 
\right) |v^F_{{\bf k},{\bf q}}(\infty) |^{2} \delta ( \omega\!-\!\mu \!+\! \varepsilon^*_{{\bf q}
\!-\!{\bf k}} \!-\!E^*_{\bf q}), \label{spectral}
\end{eqnarray}
is illustrated in Fig. 2  for T = 0.007 ($< T^*=0.016$), which lies in the pseudogap phase. 
We choose a ${\bf k}$ traversing the arc in the Brillouin zone at ${\bf k}_F(\phi)$,
in a characteristic region around $\phi = \phi_c=15^o$, 
where  the T independent gap for $\phi \leq \phi_c$ changes over into a T dependent gap in the 
single-particle density of states for values of $\phi \geq \phi_c$ (see Fig 3). $\phi_c$ 
signals the separation between localized and delocalized, respectively  bosonic and fermionic, features
in the Brillouin zone. 

For  ${\bf k}$ vectors below  ${\bf k}_F(\phi)$, $A^F({\bf k},\omega)$ exhibits 
(i) low energy ($\leq \mu$) delocalized  single-particle excitations (the first
term in eq. 8), which follow essentially the dispersion
$\varepsilon^*_{\bf k} \simeq \varepsilon_{\bf k}$ and (ii) a high energy 
($\geq \mu$) broadened upper Bogoliubov like branch. For $k \rightarrow 0$ 
that latter merges into the time reversed spectrum $-\varepsilon_{\bf k}$.  
For wave vectors  ${\bf k}$ above ${\bf k}_F(\phi)$, 
$A^F({\bf k},\omega)$ shows simultaneously two features: 
(i) low frequency diffusively propagating Bogoliubov modes and (ii) high frequency 
single-particle excitations with a dispersion given by 
$\varepsilon^*_{\bf k} \simeq \varepsilon_{\bf k}$ and  moving in a cloud of 
bosonic two-particle excitations in form  of bonding and antibonding states, 
seen by the wings on either side of the coherent part (the first term in Equ. 8)
of those excitations. These low and high frequency excitations for a given 
wave-vector characterize the low and high frequency response of one and the
same phenomenon, with the latter testing the internal degrees of 
freedom of the collective diffusively propagating Bogoliubov like modes. 
These internal degrees of freedom are images of localized bonding and 
anti-bonding states, such as given by the Green's function in the 
atomic limit ($t, t' = 0$) \cite{Domanski-1998,Domanski-2003c}, 
$G^F_{at}(i \omega_n)=1/[G^F_{at}(i \omega_n)^{-1} - \Sigma^F_{at}(i \omega_n)]$ 
with the selfenergy 
\begin{eqnarray}
\Sigma^F_{at}(i \omega_n) = {(1-Z) \; g^{2} \;(i \omega_n + \mu) \over 
[(i \omega_n + \mu)(i \omega_n - 2 \Delta + \mu) - Zg^2]},
\end{eqnarray}
\begin{figure}[t]
  \begin{center}
    \includegraphics*[width=8.5cm,angle=0]{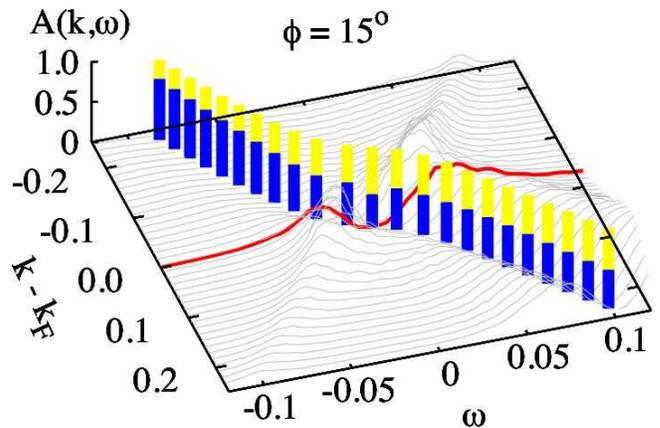}
    \caption{(Color online) $A({\bf k},\omega)$ at $T= 0.007$ ($< T^* = 0.016$) as a function 
    of $|{\bf k}|$ (in units of the inverse lattice vector) near ${\bf k}_F$ (red line), 
    corresponding to $\phi = 15^o$, orthogonally crossing the Fermi arc. The spectral weight of
    the coherent and  incoherent contributions are indicated by blue, respectively yellow bars.}
\label{fig2}
\end{center}
\end{figure}
\begin{figure}[b]
  \begin{center}
    \includegraphics*[width=8.5cm,angle=0]{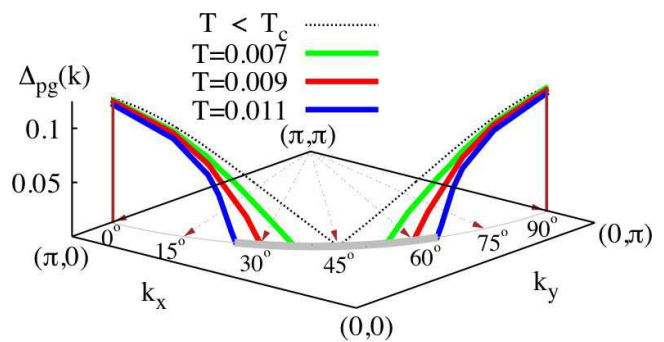}
    \caption{(Color online) Variation of the pseudogap for different k vectors, 
    orthogonally crossing the arc, given by angles $\phi$.} 
\label{fig3}
\end{center}
\end{figure}
which differs qualitatively from any BCS like structure of Cooperons, because taking into 
account their intrinsic single-particle localized internal degrees of freedom. 
$Z \simeq 2/[3+ \cosh (g/k_{B}T)]$ (for our choice of parameters) denotes the 
spectral weight of non-bonding delocalized charge carriers, described by 
$G_0^F(i \omega_n)= 1/(i \omega_n - \mu)$. 

The pseudogap in the  density of states, 
$\rho(\omega) = (1/N) \sum_{{\bf k}}A^F({\bf k}, \omega)$, which  opens up at 
some $T = T^*$   at ${\bf k}_F(\phi)$ has a size $\Delta_{pg}(\phi)$. It
is determined by the distance between the peaks either side  of 
$\varepsilon^*_{{\bf k}_F(\phi)}$, when upon lowering T the deviation from 
the bare density of state,
$\rho^0(\omega) = (1/N) \sum_{\bf k}\delta(\omega - \varepsilon_{\bf k})$ becomes 
noticeable. We take as a criterion a reduction to $90 \%$ of $\rho^0(\omega=0)$. 
The sharp peak in $A^F({\bf k}_F, \omega)$ in Fig. 2, arising from the coherent 
part of this spectrum, is a consequence of having neglected the residual 
fermion-fermion interaction $U^{F-F}_{{\bf p},{\bf k},{\bf q}}$, eq. 4. The effect
of this interaction is to broaden this delta function like peak, as we know 
from previous studies using different approaches \cite{Ranninger-1996,Robin-1998}. 
To describe this effect within the present flow equation approach, requires a 
fully self-consistent treatment of the diagonal part of the renormalized 
fermions given by 
$\sum_{{\bf k}\sigma}(\varepsilon^*_{\bf k}-\mu)c^{\dagger}_{{\bf k}\sigma}c_{{\bf k}\sigma}$ and the residual fermion-fermion interaction $H^{F-F}_{BFM}$ - 
an issue, which will be treated in some future study.

The appearance of the pseudogap is associated with a reduction of the spectral
weight of this coherent contribution (given by the height of the blue bars in 
Fig. 2). We illustrate in Fig. 3 the variation of $\Delta_{pg}(\phi)$ for 
different T. Close to the anti-nodal point - the localized and bosonic 
dominated regime - it  is relatively T independent. But approaching the nodal 
point, it  abruptly drops to zero, even though $g_{{\bf k},{\bf q}}$ is still 
finite. Although reminiscent of BCS like superconducting correlations (without 
any pseudogap) for  $60^o \geq \phi \geq 30^o$, the momentum dependence  of 
the gap in the superconducting phase is T dependent. This, clearly is a not a BCS  
mean-field type behavior \cite{Lee-2007}. The reason behind the change-over from an 
essentially  T independent gap for $\phi \leq \phi_c$ and a T dependent gap 
for $\phi \geq \phi_c$ is the following: As $\phi$ decreases, the size of 
the pseudogap increases and at the same time its 
position in the Brillouin zone at some ${\bf k}_F(\phi)$ diminishes 
until it reaches the bottom of $\varepsilon^*_{\bf k}$.
(see Fig. 2 in Ref. 37). At this point, itinerant fermionic charge 
carriers disappear in that part for the Brillouin zone, having  been converted 
into bosonic fermion pairs. The accumulation of such bosonic charge carriers near 
the anti-nodal point is a direct consequence of the anisotropic boson-femion 
exchange coupling and d-wave pairing in those cuprates. Since the excitation 
energies (size of the pseudogap) characterizing such entities are determined 
by purely local effects, they are relatively temperature as well as doping 
independent for $\phi \leq \phi_c$. Doping dependent however is the
value $\phi=\phi_c$ of the cross-over to itinerant charge carriers, as confirmed
in ARPES experiments \cite{Lee-2007}.

In order to visualize the accumulation of bosonic charge cariers near the 
anti-nodal points let us investigate how the fermionic charge carriers in 
the various regions near the arc in the Brillouin zone get converted into 
diffusively propagating bound pairs of them. To do that we evaluate the 
renormalized Bose spectral function, 
\begin{eqnarray}
A^{B}({\bf q},\omega) &=& - \frac{1}{\pi} \; \mbox{Im}  \; \int_0^{\beta} d\tau 
G^{B}({\bf q},\tau)e^{(\omega +i0^+)\tau} \nonumber \\
G^{B}({\bf q},\tau) &=& \langle \langle b_{\bf q} (\tau) ; 
b_{\bf q}^{\dagger}  \rangle \rangle_H ,
\end{eqnarray}
for which we had previously derived the corresponding  renormalization flow 
equations \cite{Domanski-2004}. It results in renormalized boson operators
\begin{eqnarray}
b_{\bf q}(\ell)= u^B_{\bf q}(\ell)  b_{\bf q} + 
\frac{1}{\sqrt{N}} \sum_{\bf k} v^B_{{\bf q},{\bf k}}(\ell)
c_{{\bf k}\downarrow} c_{{\bf q}-{\bf k}\uparrow},
\label{b_Ansatz}
\end{eqnarray}
with $b_{\bf q}^{\dagger}(\ell)= (b_{\bf q}(\ell))^{\dagger}$,  which ultimately leads to the renormalized Boson spectral function given by
\begin{eqnarray}
A^{B}({\bf q},\omega)  =  | u^B_{\bf q}(\infty)|^{2} 
\delta \left( \omega - E^*_{\bf q} \right)\nonumber \qquad  \\
 + \frac{1}{N} \sum_{\bf k} f_{{\bf k},{\bf q}-{\bf k}}
| v^B_{{\bf q},{\bf k}} (\infty) |^{2} 
\delta \left( \omega - \varepsilon^*_{\bf k}
- \varepsilon^*_{{\bf q}-{\bf k}} \right). 
\end{eqnarray}
\begin{figure}[b]
  \begin{center}
    \includegraphics*[width=8.5cm,angle=0]{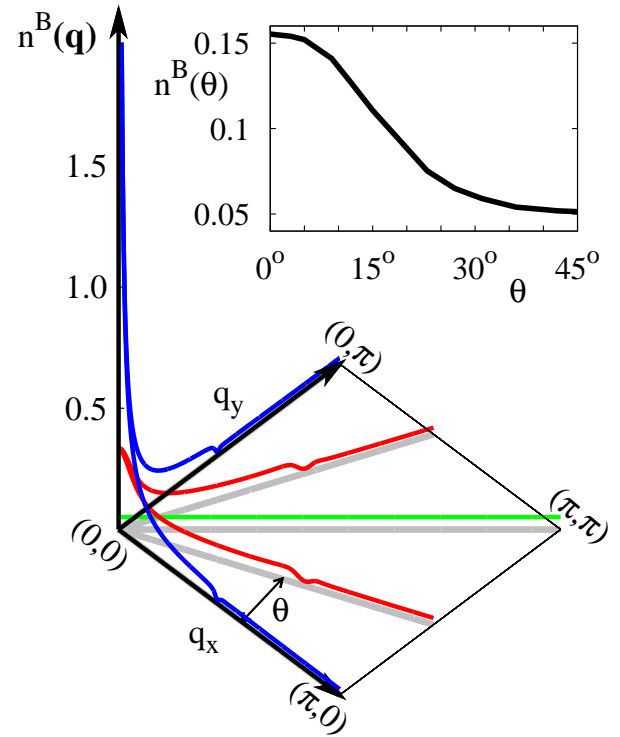}
    \caption{(Color online)  Variation of the number of paired fermions 
    as a function of wavevectors ${\bf q}$ along different directions in the Brillouin 
    zone given by the angle $\theta$. The variation of the total number of such pairs
    is illustrated in the inset.} 
\label{fig4}
\end{center}
\end{figure}
\begin{figure}[b]
  \begin{center}
    \includegraphics*[width=3.5in,angle=0]{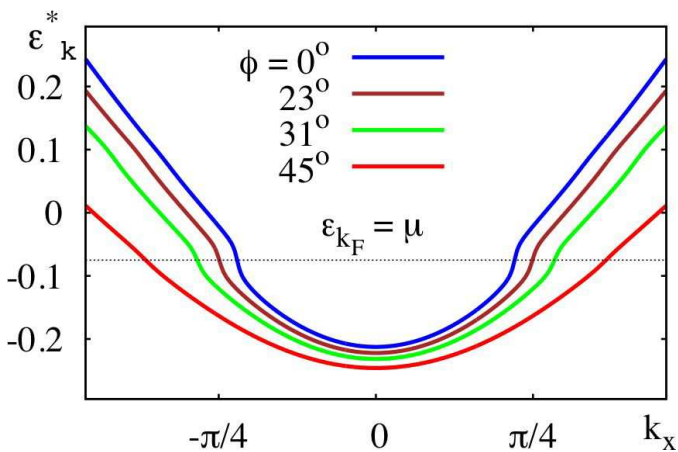}
    \caption{(Color online)  Variation of the renormalized single-particle 
    dispersion for wave-vectors $\bf k$ orthogonally crossing the arcs along 
    different directions in the Brillouin zone, characterized by the angle $\phi=arcsin(k_x/|{\bf k})|$.
    The reduction in the value of $|{\bf k}_F(\phi)|$ upon
    approaching the anti-nodal regime indicates an emptying out of the fermionic 
    single-particle excitations in favor of an increase in their paired states.} 
\label{fig5}
\end{center}
\end{figure}
The corresponding number of such bosonic charge carriers is given by
$n^B(q_x,q_y) = \int d \omega A^{B}({\bf q},\omega) [exp(\omega/k_BT) -1]^{-1}$. We 
 plot it for a series of ${\bf q}$ vectors in Fig. 4 for T=0.007, which sample the anisotropy of the CuO$_2$ electronic structure, where $\theta$ indicates the azimuthal angle in this plane. Notice that along the nodal  direction the number of bosons is independent on $|{\bf q}|$, because of the absence of any boson-fermion 
coupling. As one approaches the direction linking the center of the zone with the anti-nodal points, the 
exchange coupling steadily increases and consequently the intrinsically localized bosons acquire 
itinerancy and gather in a region of long wavelength. Those bosons have internal structure of two 
fermions with opposite momenta centered around ${\bf k}_F(\phi)$. In the inset of Fig. 4 we illustrate the 
total number of such bosons along the various ${\bf q}$ vectors and notice the relative increase, respectively 
decrease compared to their average value $0.075$, depending on whether we are sampling  the  nodal  or the  
anti-nodal  directions. The accumulation of fermions getting converted into fermion pairs 
in certain parts of the Brillouin zone, close the anti-nodal points, has its counter part in the 
diminishing density of single-particle fermionic 
excitations in the same regions. We illustrate that in Fig. 5, where we plot the variation of the coherent 
part of the single-particle dispersion, given by $\varepsilon^*_{\bf k}$ around ${\bf k}_F(\phi)$. We 
notice that with diminishing $\phi$, approaching the anti-nodal points, the corresponding value of 
${\bf k}_F(\phi)$ diminishes. This announces a shrinking of the Fermi sea, causing an emptying out of 
single-particle states and 
consequently an increase of bound fermion pairs. This feature had previously been observed in connection 
with the transition between the superconducting state of phase correlated bonding pairs and the insulating
state of such phase uncorrelated bonding pairs \cite{Stauber-2007}.

\section{Summary and Outlook}
Our scenario for the cuprate superconductivity is based on resonant pairing, 
induced by local dynamical lattice instabilities upon hole doping. It makes use 
of the fact that such systems are prone to a segregation of globally 
homogeneous crystal structures into small nano-size pairing domains. This 
breaks locally the translational as well as rotational symmetry by randomly 
orienting uni-directional Cu-O-Cu molecular bonds in different directions. As 
a result, the fermionic charge carriers acquire single-particle spectral features
which comprise simultaneously: (i) quasi localized states, where they are 
momentarily trapped in form of bound pairs in polaronic charge fluctuating 
local domains and (ii) delocalized states on a sublattice in which those
polaronic domains are embedded.

Due to the d-wave pairing, which in our case is encoded in 
the anisotropic Boson-Fermion exchange coupling $g_{{\bf k},{\bf q}}$, the 
spectral properties of the single-particle excitations exhibit a pseudogap 
with the following features: As  we move on a constant energy line in the
Brillouin zone, corresponding to the chemical potential (where such an arc 
determines the Fermi surface, whenever it exists), $|g_{{\bf k},{\bf q}}|$ 
diminishes as we go from the anti-nodal ($\phi \simeq 0$) to the nodal region 
($\phi \simeq \pi/4$). Concomitantly  the size of the pseudogap, 
$\Delta_{pg}$, decreases. For $0 < \phi <\phi_c$, with $\phi_c \simeq 15^o$ for 
our choice of parameters, it remains relatively unaffected by changes
in temperature T. On the contrary, for $\phi_c < \phi < \pi/4$, $\Delta_{pg}$ 
becomes  strongly T dependent. For low T, it tends to zero gradually as
one approaches $\phi = \pi/4$. With increasing T, it tends to zero at
increasingly larger  values of $\phi$, (see Fig.3), as observed 
experimentally \cite{Kohsaka-2008}. This suggest that:

(i) the pseudogap in a finite region  ($0 < \phi < \phi_c$) around the 
anti-nodal  point is controlled by predominately local pairing  
(via intra-{\it bonding-pair} correlations), which is independent on doping 
and largely unaffected by superconducting phase fluctuations. 

(ii)  the pseudogap  in a finite region ($\phi_c < \phi < \pi/4$) around 
the nodal point is controlled by both, local intra-{\it bonding-pair} as well 
as  non-local superconducting inter-{\it bonding-pair} correlations, which  
are sensitive to  phase fluctuations and cause the dependence of $\Delta_{pg}$ on 
T as well as on doping.

The diffusively propagating low energy Bogoliubov like excitations around 
${\bf k}_F$, which trace out the pseudogap, are a hallmark of the 
single-particle spectral features of such resonant pairing systems and which
exist even near the anti-nodal points. In contrast to a BCS scenario, here, 
their appearence above T$_c$ does not require a phase coherence of 
the bosonic bound fermion pairs. Such Bogoliubov like modes 
nucleate from local  intra-{\it bonding-pair} correlations 
between pairs of itinerant fermions  and localized fermion 
pairs \cite{Domanski-1998,Domanski-2003c} on local molecular clusters, 
such as discussed here. They are a signature of a prevailing glassy Bose 
metallic behavior prone to transit into a superconducting state of phase 
correlated such bosonic intra-{\it bonding-pairs}. The momentum dependence of 
those two-particle excitations, shows a strong tendency 
toward condensation (see Fig. 4), which tracks the anisotropic behavior of the gap. 
Provided the Boson-Fermion exchange coupling is not too big, such bosonic 
pairs forming near the anti-nodal points, will dominate the 
superconductivity, against a widespread opinion that they should be localized 
there. For sufficiently large g, they of course will be localized. This is a topic which 
will require further investigations, dealing with the superconductor to insulator 
(Bose glass) transition with reduced hole doping. 
The internal structure of those diffusively propagating Cooperons, consisting 
of selftrapped fermions, is manifest in 
their single-particle excitations above the chemical potential. It reflects 
their atomic localized nature, where two-particle localized bonding and 
anti-bonding satellites trail the dispersion of their delocalized coherent 
contributions \cite{Domanski-1998,Domanski-2003c}. The low energy diffusive 
collective Bogoliubov excitations  and the high energy single-particle 
excitations are two different manifestations of the same entity.
Whether there is a sharp border line for the onset of the high energy 
localized features in the Brillouin zone, as suggested by a socalled  
doping independent "extinction line" \cite{Kohsaka-2008,Hanaguri-2008},
will have to be checked in future for the present scenario. 

Let us conclude this study with some remarks on the kind of doping 
mechanism we can envisage in the cuprate high T$_c$ compounds.  For low hole 
doping it can be understood in terms of a doped Mott insulator and 
an antiferromagnetic ground state, transiting into a spin singlet liquid 
glassy state with increased doping. For the remaining doping regime, 
approaching the optimal and overdoped regions, it remains largely an open problem 
to be resolved. Experimentally one finds a singular universal optimal 
doping rate n$_h^{opt} = 0.16$ holes/Cu atom, where T$_c$ reaches its 
maximum together with a maximal volume fraction of the Meissner effect and 
a Hall number becoming sharply peaked \cite{Balakirev-2003}. In scenarios, 
like the present one, based on inter-related amplitude and 
phase fluctuations, optimal doping also characterizes the 
region where the energies of the superconducting phase stiffness and that of the 
pairing coincide \cite{Emery-Kivelson-1995}. These doping dependent 
electronic features are accompanied by a reduction of the buckling of the 
CuO$_2$ planes \cite{Roehler-2004}, which characterizes the low doped 
insulating phase. Pressure tuned electronic transitions, testing electronic 
and lattice features at the same time \cite{Cuk-2008}, point to a critical 
pressure which can be identified with the critical doping rate n$_h^{opt}$. 
The universal value of $n_h^{opt}$=0.16, occures for any optimally doped system, 
whatever the chemical structure of the doping blocks might be. This suggests 
that, upon approaching optimal doping, the electronic and lattice degrees of 
freedom must get strongly locked together \cite{Roehler-2009} and by doing 
so increase the stability of these intrinsically metastable materials. And 
indeed, upon trying to force extra holes into such systems  by overdoping 
$n_h > n_h^{opt}$, they segregate into different crystalline 
phases \cite{Martovitski-2007}, with superconducting components composed 
of underdoped and optimally doped regions. Understanding the doping 
dependence of the cuprates thus becomes tantamount to understanding the 
structural stability of those system. It necessarily must involve correlated 
macroscopic features \cite{Sharma-1996,Mook-1990} of charge and lattice 
deformations, such that  precisely at optimal doping they optimally and 
constructively interfere with each other. 

Transposing these experimental facts on the scenario discussed in this paper, 
the fluctuating local domains in the CuO$_2$ planes get increasingly  
coherently locked together as hole doping increases. This results in a 
decrease of spatial phase fluctuations of the bosonic resonantly bound fermion 
pairs driven by locally fluctuating lattice structures, while at the same time  
their conjugate amplitude fluctuations increase. As a consequence T$_c$ 
increases and  T$^*$ decreases. Previous studies \cite{Ranninger-2003,Cuoco-2004} 
on the interplay between amplitude and phase fluctuations bare that out. 

According to the presently available experimental facts (Ref. 13,34,35,53-55), 
the chemical doping mechanism, which imposes itself in the cuprates (following 
our scenario), converts part of the itinerant electrons into polaronically driven resonating pairs, predominantly in certain regions of the Brillouin zone (see Fig. 4) near the anti-nodal points. It  manifests itself in 
the opening of a pseudogap, which nucleates at  the socalled hot-spots, where 
the local Boson-Fermion exchange coupling g is maximal. The 
self-regulating redistribution of itinerant charge carriers and bosonic bound 
pairs of them on the  arcs in the Brillouin zone, is an intrinsic rather 
than an extrinsic \cite{Perrali-2000} feature of the scenario presented here. 
It originates from strong electron-lattice coupling, in a system with a 
highly anisotropic  electronic dispersion and coupling to local lattice 
modes, evidenced in the  anisotropic isotope dependent pseudogap and responsible 
for the local symmetry breaking of those systems. Given this experimental
situation, we conjecture that hole  doping primarily will replace the 
buckled Cu$^{II}$-O-Cu$^{II}$ bonds by unbuckled Cu$^{III}$-O-Cu$^{III}$ ones, 
whose density n$_B$ will be roughly given by $\frac{1}{2}$n$_h$, n$_h$ denoting
the concentration of  chemically doped holes. Doping a single hole into the 
basic cluster of 
our segregated CuO$_2$ planes means a doping rate of 1/8= 0.125 per Cu ion. 
This  is very close to the critical doping rate, which changes the insulating 
glassy phase into the superconducting one. Doping a hole into the trapping 
centers breaks a Cu$^{II}$-O-Cu$^{II}$ bond. Since this is not compatible with  
the basic square planar CuO$_4$ structure in the CuO$_2$ planes, doping  will 
trigger a charge  transfer between the trapping centers and the surrounding 
four-site rings, either by transferring an electron from the ring to the trapping 
center and re-establish the  stable square planar Cu$^{II}$-O-Cu$^{II}$ bond, or  
by transferring  an electron from that hole doped bond into the ring and leave behind 
a stable square planar Cu$^{III}$-O-Cu$^{III}$ bond. 
Both of these processes act together to ensure the overall crystalline
stability in systems with intrinsic local dynamically correlated 
charge-lattice fluctuations and  thus result in
resonant pairing of the itinerant electrons on the ring. The 
end-effect of this is a transfer of a fraction n$_F^h$ of the electrons 
on the ring into the pairing centers, where they form pairs on a finite time 
scale with a concentration n$_B$ = $\frac{1}{2}$n$_F^h$. This simultaneously 
implies a shrinking of the Fermi surface. n$_{tot}$ = n$_F$ + 2n$_B$ in this doping 
procedure remains unaltered  i.e., equal to unity as it is in the undoped case. 
The effect of hole doping is hence to change the relative concentration of itinerant electrons with respect to the concentration of partly bound pairs of them.

A multitude of different experimental results discussed here have been shown 
to be compatible with the resonant pairing scenario. Qualitatively different 
from any BCS  pairing scenario, here the itinerant delocalized Bogoliubov
excitations coexist with localized single-particle ones which  
are selftrapped inside of them. Concerning the origin of this resonant
pairing in the cuprates, which could be electronic \cite{Altman-2002}, as 
well as polaronic, the recently observed breakdown of their homogeneous
crystal structures into translational/rotational symmetry  broken local 
clusters \cite{Kohsaka-2008}, gives us confidence that dynamical lattice 
deformations should play a determinant role  in the superconducting state 
of high T$_c$ compounds.

\section{Acknowledgement}

We thank Juergen Roehler for constructive remarks concerning this work and its 
presentation.


\end{document}